\documentclass[sigconf]{acmart}
\AtBeginDocument{
  \providecommand\BibTeX{{
    \normalfont B\kern-0.5em{\scshape i\kern-0.25em b}\kern-0.8em\TeX}}}

\setcopyright{none}
\renewcommand\footnotetextcopyrightpermission[1]{}
\setcopyright{acmcopyright}
\copyrightyear{2023}
\acmYear{2023}
\acmDOI{XXXXXXX.XXXXXXX}

\acmPrice{15.00}
\acmISBN{978-1-4503-XXXX-X/18/06}

\usepackage{amsmath,amsfonts}
\usepackage{algorithm}
\usepackage{graphicx}
\usepackage{textcomp}
\usepackage{xcolor}
\usepackage{hyperref}
\usepackage{xspace}
\usepackage{multirow}
\usepackage{tcolorbox}
\usepackage{color, colortbl}
\usepackage{listings}
\usepackage{nicefrac}
\usepackage{tabularx}

\newcommand{\etal}{\emph{et al}.\@\xspace}
\newcommand*{\eg}{\emph{e.g.},\@\xspace}
\newcommand*{\ie}{\emph{i.e.},\@\xspace}
\newcommand*{\aka}{\emph{a.k.a.}\@\xspace}

\definecolor{codegreen}{rgb}{0,0.6,0}
\definecolor{codegray}{rgb}{0.5,0.5,0.5}
\definecolor{codepurple}{rgb}{0.58,0,0.82}
\definecolor{backcolour}{rgb}{0.95,0.95,0.95}

\begin{document}

\title[The Importance of Discerning Flaky from Fault-triggering Test Failures]{The Importance of Discerning Flaky from Fault-triggering Test Failures: A Case Study on the Chromium CI\\ \ }


\author{Guillaume Haben}
\affiliation{%
  \institution{SnT, University of Luxembourg}
  \country{Luxembourg}
}
\email{guillaume.haben@uni.lu}

\author{Sarra Habchi}
\affiliation{%
  \institution{Ubisoft}
  \country{Canada}
}
\email{sarra.habchi@ubisoft.com}

\author{Mike Papadakis}
\affiliation{%
  \institution{SnT, University of Luxembourg}
  \country{Luxembourg}
}
\email{michail.papadakis@uni.lu}

\author{Maxime Cordy}
\affiliation{%
  \institution{SnT, University of Luxembourg}
  \country{Luxembourg}
}
\email{maxime.cordy@uni.lu}

\author{Yves Le Traon}
\affiliation{%
  \institution{SnT, University of Luxembourg}
  \country{Luxembourg}
}
\email{yves.letraon@uni.lu}

\renewcommand{\shortauthors}{Haben, et al.}

\begin{abstract}
Flaky tests are tests that pass and fail on different executions of the same version of a program under test. They waste valuable developer time by making developers investigate false alerts (flaky test failures). 
To deal with this problem, many prediction methods that identify flaky tests have been proposed. While promising, the actual utility of these methods remains unclear since they have not been evaluated within a continuous integration (CI) process. In particular, it remains unclear what is the impact of missed faults, \ie the consideration of fault-triggering test failures as flaky, at different CI cycles. To fill this gap, we apply state-of-the-art flakiness prediction methods at the Chromium CI and check their performance. Perhaps surprisingly, we find that, despite the high precision (99.2\%) of the methods, their application leads to numerous faults missed, approximately 76.2\% of all regression faults.
To explain this result, we analyse the fault-triggering failures and show that flaky tests have a  strong fault-revealing capability, \ie, they reveal more than \nicefrac{1}{3} of all regression faults, indicating an inherent limitation of all methods focusing on identifying flaky tests, instead of flaky test failures. Going a step further, we build failure-focused prediction methods and optimize them by considering new features. Interestingly, we find that these methods perform better than the test-focused ones, with an MCC increasing from 0.20 to 0.42. Overall, our findings imply that on the one hand future research should focus on predicting flaky test failures instead of flaky tests and the need for adopting more thorough experimental methodologies when evaluating flakiness prediction methods, on the other.

\end{abstract}






\maketitle
\pagestyle{plain}

\section{Introduction}
\label{sec:introduction}

Continuous Integration (CI) is a software engineering process that allows developers to frequently merge their changes in a shared repository~\cite{CI}. To ensure a fast and efficient collaboration, the CI automates parts, if not all, of the development life cycle. Regression testing is an important step of this cycle as it ensures that new changes do not break existing features.  Test suites are executed for every commit and test results signal whether changes should be integrated into the codebase or not.

Tests are an essential part of the CI as they prevent faults from interring the codebase, and they ensure smooth code integration and overall good software function. Unfortunately, some tests, named flaky tests, exhibit a non-deterministic behaviour as they both pass and fail for the same version of the codebase. When flaky tests fail, they send false alerts to developers about the state of their applications and the integration of their changes. 

Indeed, developers spend time and effort investigating flaky failures, as they can be difficult to reproduce, only to discover that they are false alerts ~\cite{Eck2019}. These false alerts occur frequently in open-source and industrial projects~\cite{Bell2018,Kowalczyk2020,Lam2019iDFlakies,LeongSPTM19} and make developers lose not only time but also their trust in the test signal. This trust issue in turn introduces the risk of ignoring fault-triggering test failures. This way, false alerts defy the purpose of software testing and hinder the flow of the CI.

To deal with test flakiness, many techniques aiming at detecting flaky tests have been introduced. A basic approach is to rerun tests multiple times and observe their outcomes. While to some extent effective, test reruns are extremely expensive \cite{LeongSPTM19,Bell2018} and unsafe. To this end, researchers have proposed several approaches relying on static (the test code) \cite{camara2021use,Pinto2020,fatima2021flakify,King2018, LeongSPTM19} or dynamic (test executions) \cite{Bell2018,Lam2019iDFlakies,ziftci2020flake} information (or both) \cite{alshammari2021flakeflagger} to predict whether a given test is flaky. 

Among the many flakiness prediction methods, the vocabulary-based ones~\cite{Haben2021,Pinto2020,Camara2021VocabExtendedReplication,Bertolino2020,olewickiBrown} are the most popular \cite{ParryKHM22}. They rely on machine learning models that predict test flakiness based on the occurrences of source code tokens of the candidate tests. Interestingly, previous research has found these approaches particularly precise, with current state-of-the-art achieving accuracy values higher than 95\% \cite{Pinto2020,fatima2021flakify,camara2021use,Camara2021VocabExtendedReplication,Haben2021}. 

At the same time, vocabulary-based approaches are static and text-based, thus, they are both portable, \ie limited to a specific language, and interpretable, \ie it is easy to share information about important keywords impacting the model's decisions. All these characteristics (precision, portability and interpretability) make vocabulary-based approaches flexible and easy to use in practice. Therefore, we decided to replicate these techniques on an industrial project (the Chromium project) and evaluated their ability to effectively support the CI process. 

Perhaps not surprisingly, we found a similar prediction performance (99.2\% precision 98.4\% recall) to the ones reported by previous studies. Surprisingly though, we noticed numerous fault-triggering failures (approximately 76\% of all fault-triggering failures) being marked as flaky by these prediction methods. This means that, in the case where the Chromium teams were to follow the recommendations of these techniques, they would have missed at least 76\% of all regression faults (considering their respective test failures as flaky) that were captured by their test suites. 

The paradox of having high flaky test prediction precision and high fault loss (fault-triggering failures considered as flaky) motivated the investigation of the fault-triggering failures. To this end, we made the following three findings: 
\begin{itemize}
    \item Flaky tests have a strong fault-triggering capability, \ie more than \nicefrac{1}{3} of all regression faults are triggered by a test that exhibits flaky behaviour at some point in time. This means that methods aiming at detecting flaky tests, inevitably flag as flaky fault-triggering test failures made by these tests. This indicates an inherent limitation of all methods focusing on identifying flaky tests, a fact largely ignored in previous studies.
    
    \item Many fault-triggering failures have characteristics similar to flaky ones and thus, are mistakenly flagged as flaky. While such mistakes are expected, given the predictive nature of the approaches, these are expected to be low given the 99.2\% prediction precision we have. However, in our data, this set of false classifications represents 56.2\% of all fault-triggering failures. 

    \item The majority of the flaky tests fail frequently (87.9\% of the flaky tests has also flaked in the past), making prediction methods mark them as flaky based on the test history, instead of the characteristics of their failures. In particular, a dummy method that classifies as flaky any test that flaked at least once in the past achieves precision and recall values of 99.8\% and 87.8\% when predicting flaky test failures.
\end{itemize}

The above findings motivate the need for techniques focusing on flaky test failures, instead of flaky tests, \ie discriminating between fault-triggering and flaky test failures, an essential problem that has largely been ignored by previous research \cite{Parry2021}. Therefore, we adapt the vocabulary-based methods for failure-focused predictions and check their performance. Although they miss fewer fault-triggering test failures than the test-focused methods (20.3\% FPR compared to 76.2\% FPR), their ability to detect fault-triggering failures remains poor, MCC value of 0.25. 

With the hope of improving the methods' performance, we augment their feature set with dynamic features related to (flaky) test executions (\eg run duration, test outcome and historical flake rate) that we find useful for flakiness diagnosis. We achieve a better --but still not acceptable performance -- MCC value of 0.42, indicating an improvement to detect fault-triggering failures when considering dynamic features. 

Overall, our study demonstrates the need for methods that can effectively predict flaky test failures (instead of flaky tests) and the need for adopting more thorough experimental methodologies, reflecting the needs of the domain of the actual practice (not just classification metrics), when evaluating flakiness predictions. 

In summary, the contributions of our paper are:
\begin{itemize}
    \item We present a large empirical study on flakiness prediction, based on the Chromium project -- one of the biggest open-source industrial projects -- involving 10,000 builds, more than 200,000 unique tests and 1.8 million test failures. Our study is the first to study the suitability of applying flakiness prediction into a CI pipeline by focusing on the potential losses that they introduce: missed fault-triggering failures.  
    
    \item We provide empirical evidence that flaky test prediction methods, despite being very precise, are practically non-actionable since they flag as flaky a majority, approximately 76.2\%, of all fault-triggering failures (56.2\% due to model misclassifications and 20\% due to correct classification of flaky tests that reveal faults).

    \item We provide empirical evidence that flaky tests have strong fault-triggering capabilities, indicating an inherent limitation of existing methods. At the same time, our results motivate the need for failure-focused prediction techniques. Unfortunately, we also show that existing vocabulary-based methods are insufficiently precise, calling for additional research in this area.  

    \item We investigate whether imbuing the training data with additional dynamic features might enhance the failure prediction efficacy. These results show  improvement (MCC values are up to 0.42), but indicate that more work remains to be done to develop deployable flakiness prediction for real-world systems.
   
\end{itemize}

To support future research, we share our dataset, experimental data and related code, in a replication package.\footnote{https://anonymous.4open.science/r/ChromiumFlakyFailures}

\section{Related Work}\label{section:related}

\begin{table*}[]
\begin{center}
\caption{Existing machine learning-based studies aiming at detecting test flakiness. The majority of the techniques focus on detecting flaky tests, while half of the approaches rely on vocabulary features.}
\vspace{-0.6em}
\label{table:relatedWork}
\begin{tabular}{c | c | c | c | c | c | c}
\toprule
\textbf{Study} & \textbf{Model} & \textbf{Feature category} & \textbf{Features} & \textbf{Benchmark} & \textbf{Target}& \textbf{Year}\\ \midrule
King et al.~\cite{King2018}                          & Bayesian network                    & Static \& dynamic                              & Code metrics           & Industrial                              & \textbf{Flaky tests} &  2018                        \\ 
Pinto et al.~\cite{Pinto2020}                         & Random forest                       & Static                                         & \textbf{Vocabulary}             & DeFlaker                                & \textbf{Flaky tests} &          2020                \\ 
Bertolino et al.~\cite{bertolino2020learning}                     & KNN                                 & Static                                         & \textbf{Vocabulary}             & DeFlaker                                & \textbf{Flaky tests}                     &   2020  \\ 
Haben et al.~\cite{Haben2021}                         & Random forest                       & Static                                         & \textbf{Vocabulary}             & DeFlaker                                & \textbf{Flaky tests}  &           2021             \\ 
Camara et al.~\cite{Camara2021VocabExtendedReplication}                        & Random forest                       & Static                                         & \textbf{Vocabulary}                 & iDFlakies                               & \textbf{Flaky tests}                &    2021      \\ 
Alshammari et al.~\cite{alshammari2021flakeflagger}                    & Random forest                       & Static \& dynamic                              & \begin{tabular}{@{}c@{}}Code metrics \&\\ Smells\end{tabular}  & FlakeFlagger                            & \textbf{Flaky tests}                        &  2021 \\ 
Fatima et al.~\cite{fatima2021flakify}                        & Neural Network                      & Static                                         & CodeBERT               & \begin{tabular}{@{}c@{}}FlakeFlagger \\iDFlakies\end{tabular}               & \textbf{Flaky tests}                     &   2021  \\ 
Pontillo et al.~\cite{Pontillo}                      & Logistic regression                 & Static                                         &\begin{tabular}{@{}c@{}}Code metrics \&\\ Smells\end{tabular} & iDFlakies                               &  \textbf{Flaky tests} &       2021      \\ 
Lampel et al.~\cite{lampelOrange}                        & XGBoost                             & Static \& dynamic                              &  \begin{tabular}{@{}c@{}}Job execution \\metrics \end{tabular}                       & Industrial                              & Flaky failures               &   2021     \\ 
Qin et al.~\cite{qin2022peeler}                      & Neural Network                             & Static                                         & Dependency graph           & Industrial                              & \textbf{Flaky tests}                     & 2022  \\ 
Olewicki et al.~\cite{olewickiBrown}                      & XGBoost                             & Static                                         & \textbf{Vocabulary}             & Industrial                              & Flaky builds                      & 2022  \\ 
Ackli et al.~\cite{flakicat}                      & Siamese Networks                             & Static                                         & CodeBERT             & Various                             & \textbf{Flaky tests}                     & 2022  \\ \bottomrule
\end{tabular}
\end{center}
\vspace{-0.5em}
\end{table*}

\subsection{Test Flakiness}
Flakiness is a known issue in software testing but research studies on this topic have only gained momentum in the past few years \cite{Parry2021}. Luo \etal~\cite{Luo2014} conducted the first empirical study on the root causes of flakiness. They analyzed commit fixes from 51 open-source projects and created the first taxonomy of flaky tests. Later on, several  studies on test flakiness followed using different settings. Lam \etal~\cite{Lam2019RootCausing} conducted a study on flaky tests at Microsoft to identify and understand their root causes. They presented \textit{RootFinder}, a framework that helps to debug flaky tests using logs and time differences in their passing and failing runs. 

Romano \etal~\cite{romano2021empirical} analysed User Interface (UI) tests and showcased the flakiness root causes and the conditions needed to fix them. In the same vein, Gruber \etal \cite{Gruber2021} presented a large empirical analysis of more than 20,000 Python projects. They found test order dependency and infrastructure to be among the top reasons for flakiness in those projects.

To help debug, reproduce, and comprehend the causes of flaky tests several tools have been introduced.  \textit{DeFlaker}~\cite{Bell2018} detects flaky tests across commits, without performing any reruns, by checking for inconsistencies in test outcomes with regard to code coverage. Focused on test order dependencies, \textit{iDFlakies}~\cite{Lam2019iDFlakies} detects flaky tests by rerunning test suites in various orders. 

To increase the chances of observing flakiness, Silva \etal~\cite{Silva2020} introduced \textit{Shaker}, a technique that relies on stress testing when rerunning potential flaky tests.  Another line of work aims at repairing (atuomatically) flaky tests. To this end, Shi~\etal~\cite{Shi2019iFix} proposed iFixFlakies, a framework that recommends patches for order-dependent flaky tests based on test patterns found in non-flaky tests that exhibit similar behaviour as flaky tests. 

\subsection{Flakiness Detection Methods}
While they remain scarce, the recent publication of datasets of flaky tests~\cite{Bell2018,Lam2019iDFlakies,Gruber2021} enabled new lines of work. Prediction models were suggested to differentiate flaky tests from non-flaky tests. King \etal~\cite{King2018} presented an approach that leverages Bayesian networks to classify and predict flaky tests based on code metrics. Pinto \etal~\cite{Pinto2020} used a bag of words representation of each test to train a model able to recognize flaky tests based on the vocabulary from the test code. This line of work has gained a lot of momentum lately as models achieved higher performances. Several works were carried out to replicate those studies and ensure their validity in different settings~\cite{Haben2021,Camara2021VocabExtendedReplication}.

In an industrial context, Kowalczyk \etal~\cite{Kowalczyk2020} implemented a flakiness scoring service at Apple. Their model quantifies the level of flakiness based on their historical flip rate and entropy (\ie changes in test outcomes across different revisions). Their goal was to identify and rank flaky tests to monitor and detect trends in flakiness. They were able to reduce flakiness by 44\% with less than 1\% loss in fault detection. In our study, we also rely on test history to help with flakiness prediction.

More recently, \textit{FlakeFlagger}~\cite{alshammari2021flakeflagger} has been proposed. It builds a prediction model using an extended set of features from the code under test together with test smells. The research community continue to draw attention to this field by considering other possible features to predict flaky tests, this is the case of Peeler~\cite{qin2022peeler} for example, where the authors leveraged test dependency graphs to predict flaky tests.

Less attention has been given to flaky failures or false alerts prediction. Herzig \etal~\cite{Herzig2015} used association rules to identify false alert patterns in the specific case of system and integration tests that contain steps. They evaluated their approach on Windows 8.1 and Microsoft Dynamics AX.

Olewicki \etal investigated the possibility of leveraging vocabulary-based features on logs from failing builds to predict if failures are caused by defects in the code or by other non-deterministic issues including flaky test failures. It is interesting to note that their work focuses on builds and not tests as we do in our study.

Finally, a recent study by Lampel \etal~\cite{lampelOrange} presented an approach that automatically classifies jobs by deciding if a job failure originates from a bug in the code or from flakiness. To do so, they relied on features from job executions, \eg CPU load, and run time. As such they are only concerned with some form of flaky failures and not with the utility of detecting flaky failures in CI, instead of tests as we do in this paper. 

Table~\ref{table:relatedWork} summarizes the state-of-the-art of flakiness prediction in chronological order. By inspecting the table, we see that most of the studies focus on flaky tests, with just one focusing on flaky test failures. We also notice that static features like code metrics and test smells are often used but features based on vocabulary (\ie bag-of-words) are the most popular ones. 

Overall, in contrast to previous work, our study is the first one that investigates the utility of the prediction methods in CI, in relation to the risk of missing fault-triggering test failures. 

\section{The Chromium Project}
\label{section:chromium}

\subsection{Overview}

Started in 2008, with more than 2,000 contributors and 25 million lines of code, the Chromium web browser is one of the biggest open-source projects currently existing. Google is one of the main maintainers, but other companies and contributors are also taking part in its development.

Chromium relies on \textit{LuCI} as a CI platform~\cite{onlineChromiumGithub}.
It uses more than 900 parallelized builders, each one of them used to build Chromium with different settings (\eg different compilers, instrumented versions for memory error detection, fuzzing, etc) and to target different operating systems (\eg Android, Mac OS, Linux, and Windows). 
Each builder is responsible for a list of builds triggered by commits made to the project. If a builder is already busy, a scheduler creates a queue of commits waiting to be processed. This means that more than one change can be included in a single build execution if the development pace is faster than what the builders can process. Within a build, we find details about build properties, start and end times, status (\ie pending, success or failure), a listing of the steps and links to the logs. 

In the beginning of the project, building and testing were sequential. Builders used to compile the project and zip the results to builders responsible for tests. Testing was taking a lot of time, slowing developers' productivity and testing Chromium on several platforms was not conceivable. A swarming infrastructure was then introduced in order to scale according to the Chromium development team's productivity, to keep getting the test results as fast as possible and independently from the number of tests to run or the number of platforms to test. A fleet of 14,000  build bots runs tasks in parallel. This setup helps to run tests with low latency and handle hundreds of commits per day~\cite{TheChromiumProjects}.

\begin{figure}[ht]
\centering
\includegraphics[width=0.45\textwidth]{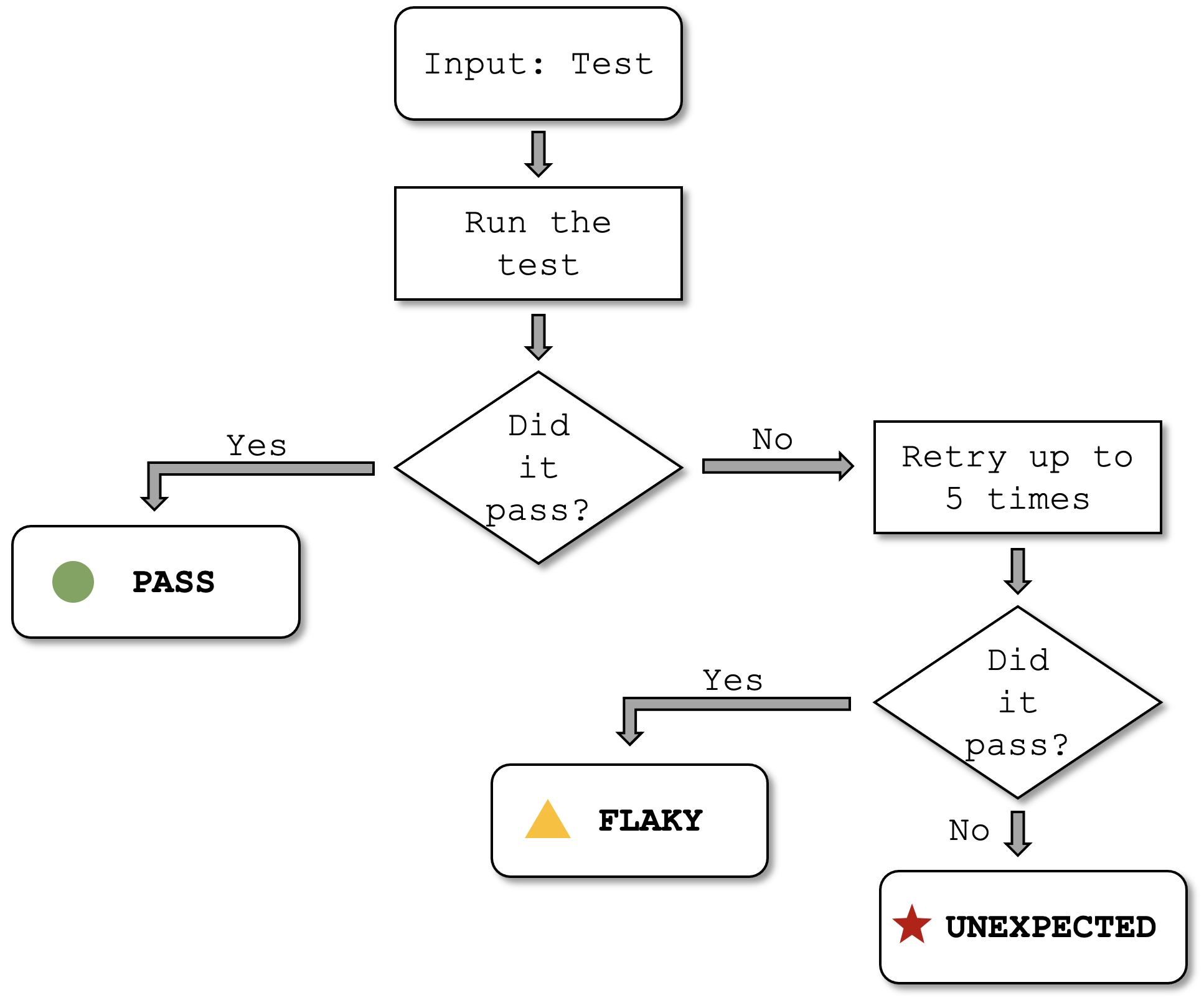}
\caption{Decision tree representing how test outcomes are determined in a build by the Chromium CI. \includegraphics[scale=0.2]{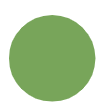} PASS depicts successful tests, \includegraphics[scale=0.2]{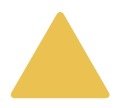} FLAKY depicts tests that passed after failing at least once, while \includegraphics[scale=0.2]{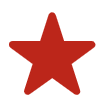} UNEXPECTED depicts tests that persistently failed.}
\label{fig:decision-tree}
\end{figure}

In this study, we focus on testers, \ie builders only responsible for running tests. They do not compile the project: when triggered, they simply extract the build from their corresponding builder and run tests on this version. At the time of writing, we found 47 testers running Chromium test suites on distinct operating systems versions. About 200,000 tests are divided into different test suites, the biggest ones being \textit{blink\_web\_tests} (testing the rendering engine) and \textit{base\_unittests} with more than 60,000 tests each.

For each build performed by any tester, we have access to information about test results. Figure~\ref{fig:decision-tree} illustrates the decision process followed by LuCI to determine a test outcome in a specific build. A test is labelled as \textit{pass} when it successfully passed after one execution. In case of a failure, LuCI automatically reruns the test up to 5 times. If all reruns fail, the test is labelled as \textit{unexpected} and will trigger a build failure. In the remaining, we will be referring to \textit{unexpected} tests as \textit{fault-revealing tests}. If a test passes after having one or more failed executions during the same build, it is labelled as \textit{flaky} and will not prevent the build from passing.

\subsection{Example of a flaky test}


\begin{figure}[ht]
\centering
\includegraphics[width=0.42\textwidth]{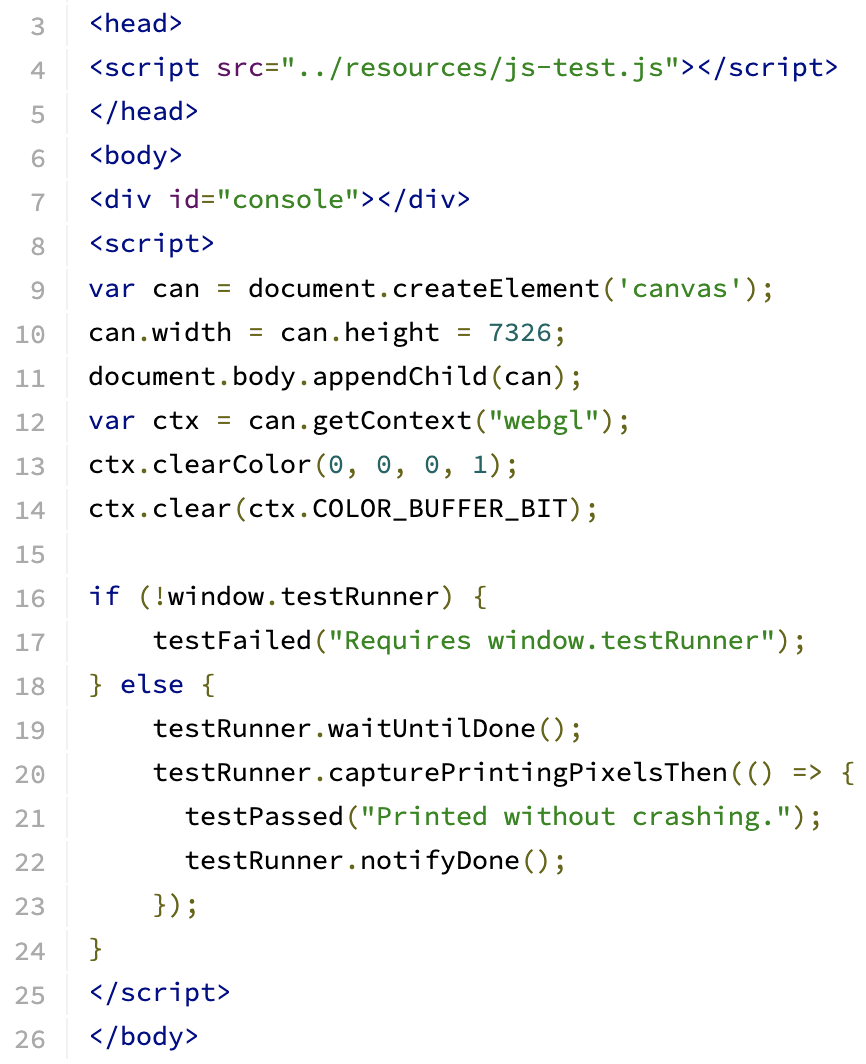}
\vspace{-1.0em}
\caption{An example of a flaky test caused by a timeout. The test consists of an HTML file \texttt{printing/webgl-oversized-printing.html}, build 119,039 of the Linux Tester. The call to \textsc{waitUntildone()} on line 19 is likely the reason for the failure.}
\label{fig:example}
\vspace{-0.5em}
\end{figure}

Figure~\ref{fig:example} shows a flaky test found in build 119,039\footnote{https://ci.chromium.org/ui/p/chromium/builders/ci/Linux\%20Tests/119039/} of the Linux Tester. This test, \texttt{printing/webgl-oversized-printing.html}, ensures that no crash happens on the main thread of the rendering process when using the system. Unfortunately, on its first execution, the test failed after 31 seconds. The run status indicates that a \textsc{TIMEOUT} happened. On the second execution, the test passed after 15 seconds and thus was labelled as flaky. In this case, an issue has been opened in Chromium's bug tracking system.\footnote{https://bugs.chromium.org/p/chromium/issues/detail?id=1393294} Developers state that \textit{"this test makes a huge memory allocation in the GPU process which intermittently causes OOM and a GPU process crash"}.

\textsc{Timeout} is a run status that intuitively leads to possible flakiness, as we can easily think of other executions of the same test being completed before reaching the time limit. In addition to this feature, we can also look for hints of flakiness in the source code. As of many UI tests in Chromium, this one is handled by \texttt{testRunner}: a test harness in charge of their automatic executions. We can see the \texttt{testRunner} making a call to \texttt{waitUntilDone()} on line 19. Vocabulary about waits is common in Chromium's web tests. This keyword, for example, could potentially be leveraged by flakiness detectors to classify tests or failures.


\section{Data}
\label{sec:dataset}

\subsection{Definitions}

Some of our definitions are slightly different from the one used by previous work since Chromium has its specific continuous integration setup. To make things clear, we define the elements we will discuss in this section.  
In the scope of \emph{a single build}, we employ the following definitions: 
\begin{itemize}
    \item \textbf{Fault-revealing test}: A test that consistently failed after reruns in the same build, revealing a regression.
    \item \textbf{Flaky test}: A test that failed once or more and then passed after reruns in the same build. 
    \item\textbf{Flaky failure}: A failure caused by a flaky test.
    \item\textbf{Fault-triggering failure}: A failure caused by a fault-revealing test.\\
\end{itemize}

\subsection{Data collection}

To perform our study, we collected test execution data from 10,000 consecutive builds completed by the Linux Tester by querying the LuCI API. This represents a period of time of about 9 months taken between March 2022 and December 2022. 

Table ~\ref{table:infoRuns} summarises the list of information extracted and computed for all tests executed in all builds. The \textsc{buildId} corresponds to the build in which tests were executed. Run duration is the execution time spent to run the test. Run status gives information about the run result (\eg passing, failing, and skipped) and run tag status returns more precise information about the result of a run depending on the type of test or test suite (\eg timeout and failure on exit). We retrieved information about the tests' source code by querying Google Git~\footnote{https://chromium.googlesource.com/chromium/src/+/HEAD/}. As builds often handle several commits, we select the revision corresponding to the head of the blame list: the one on which tests were executed. The test suite is simply the name of the test suite the test belongs to and testId is a unique identifier for a test composed of the test suite and the test name (the same test name can be present in different test suites).

All the scripts used to collect the data alongside the created dataset are available in our replication package.

\begin{table}[ht]
\begin{center}
\caption{Description of our features. Column \textit{Feature Name} specifies the identifiers used in our dataset, while Column \textit{Feature Description} details the features}
\label{table:infoRuns}
\begin{tabular}{ l | l } 
\toprule
\textbf{Feature Name} & \textbf{Feature Description} \\
\midrule
buildId & The build number associated \\
{} & with the test execution \\
\midrule
flakeRate & The flake rate of the test over the last\\  
{} & 35 commits \\
\midrule
runDuration & The time spent for this test execution \\
\midrule
runStatus & ABORT\\ {} & FAIL\\ {} & PASS\\ {} & CRASH\\ {} & SKIP \\
\midrule
\multirow[t]{9}{*}{runTagStatus} & CRASH\\ {} & PASS\\ {} & FAIL\\ {} & TIMEOUT\\ {} & SUCCESS \\
{} & FAILURE\\ {} & FAILURE\_ON\_EXIT\\ {} & NOTRUN\\ {} & SKIP\\ {} & UNKNOWN \\
\midrule
testSource & The test source code \\
\midrule
testSuite & The test suite the test belongs to \\
\midrule
testId & The test name \\
\bottomrule
\end{tabular}
\end{center}
\vspace{1.0em}
\end{table}

\begin{table*}[ht]
\begin{center}
\caption{Data collected from the Chromium CI. We used the \textit{Linux Tests} tester, with \textit{10,000} Builds mined over \textit{nine} months. We extracted Passing, Flaky and Fault-revealing tests and their associated Flaky and Fault-triggering Failures.}
\vspace{-0.5em}
\label{table:infoDataset}
\begin{tabular}{ l | c | c c | c c c | c c } 
\toprule
\multirow{2}{*}{\textbf{Tester}} & \multirow{2}{*}{\textbf{Nb of Builds}} & \multicolumn{2}{c|}{\textbf{Period of Time}} & \multicolumn{3}{c|}{\textbf{Number of Tests}} & \multicolumn{2}{c}{\textbf{Number of Failures}} \\
{} & {} & \textbf{From} & \textbf{To} & \textbf{Passing} & \textbf{Flaky} & \textbf{Fault-revealing} & \textbf{Flaky} & \textbf{Fault-triggering} \\
\midrule 
Linux Tests & 10,000 & Mar 2, 2022 & Dec 1, 2022 & 198,273 & 23,374 & 2,343  & 1,833,831 & 17,171 \\
\bottomrule
\end{tabular}
\end{center}
\end{table*}

\subsection{Computing the flake rate}
\label{sec:testHistory}

The historical sequence of test results is a valuable piece of information commonly used in software testing at scale~\cite{LeongSPTM19,Kowalczyk2020}. We analyse the history of fault-revealing tests and flaky tests by relying on their \textsf{flakeRate}.

This means that for a test $t$ failing (due to flaky or fault-triggering failure) in a build $b_{n}$, we analyse all the builds from a time window $w$ (\ie from $b_{n-w}$ to $b_{n-1})$ to calculate its flake rate as follows: \\

\noindent\begin{minipage}{\linewidth}
\begin{equation}
  flakeRate(t,n) = \frac{ \sum_{x=n-w}^{n-1} flake(t,x) } {w}
\end{equation}
\end{minipage}%
\\

where $flake(t,x) = 1$ if the test $t$ flaked in the build $b_{x}$ and 0 otherwise.
This metric allows us to understand if the flakiness history of a test can help in the flakiness prediction tasks.
The test execution history (\aka heartbeat) has been used in multiple studies (especially industrial ones~\cite{Kowalczyk2020,LeongSPTM19}) to detect flaky tests.
These studies assume that many flaky tests have distinguishable failure patterns over builds and hence can be detected by observing their history.
We check whether this assumption also holds in the case of Chromium.

\begin{figure}[!htbp]
    \vspace{-1em}
  \centering
    \includegraphics[width=0.5\textwidth]{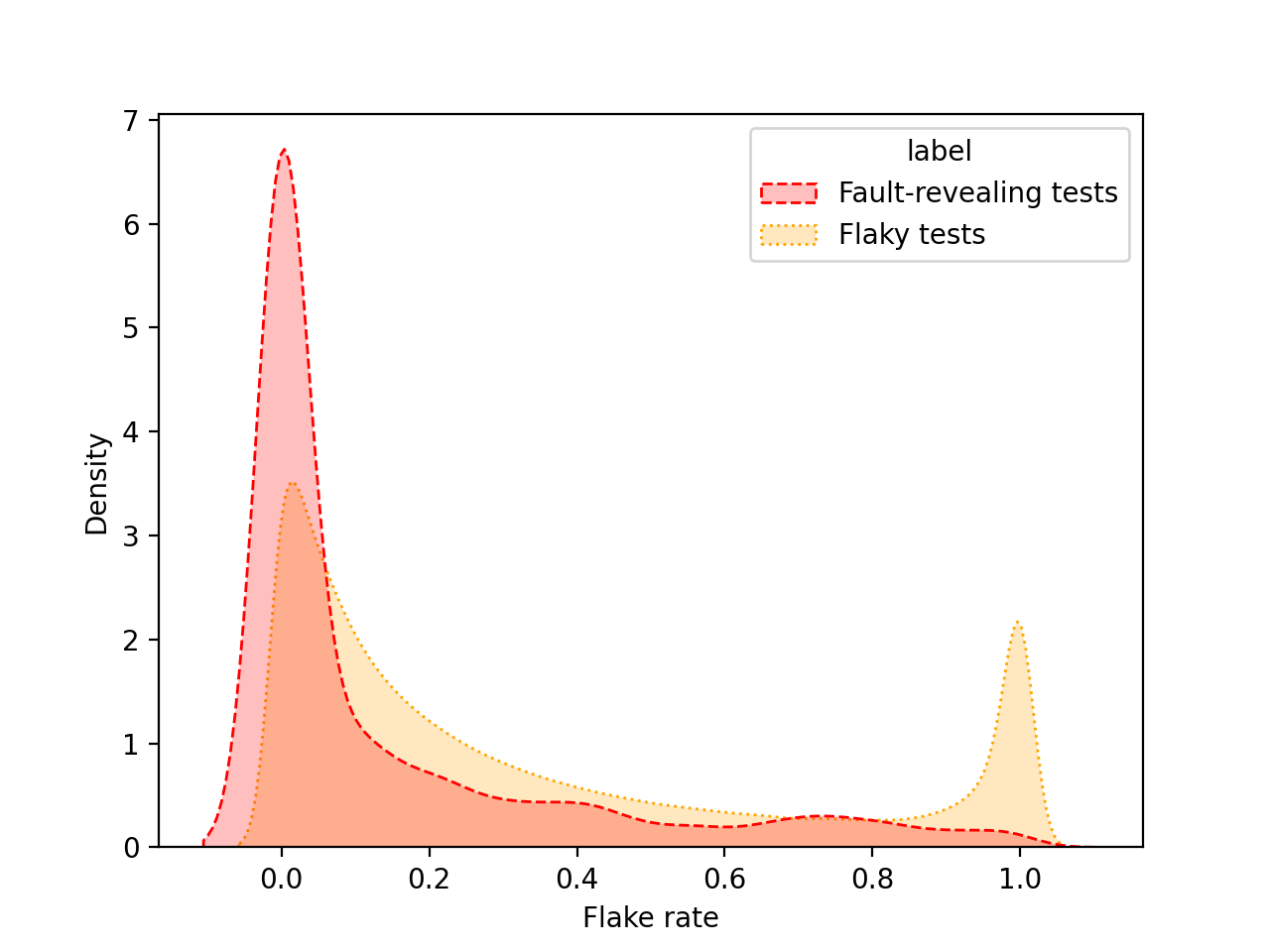}
    \vspace{-1em}
    \caption{Flake rate (\textit{x-axis}) for \textcolor{orange}{Flaky} and \textcolor{red}{Fault-revealing tests}. Density (\textit{y-axis}) is the probability density function. The area under curves integrates to one. Many flaky tests are always flaky in their previous builds. A majority of fault-revealing tests have no history of flakiness at all.}
    \label{fig:kdeRates}
\end{figure}

To illustrate the flake rate differences between flaky and fault-revealing tests, we plot the flake rate for both test categories in Figure~\ref{fig:kdeRates}. The flake rate is computed using a window of 35 builds. To set this time window, we checked the number of flaky tests having a $flakeRate() = 0$ for build windows ranging from 0 to 40 builds with a step of 5. We observed a convergence at size 35, meaning that higher numbers of builds do not provide additional information.

In the majority of cases, flaky tests have a history of flakiness: the percentage of flaky tests having a $flakeRate() > 0$ is in fact $87.9\%$. Furthermore, we see a pike for $flakeRate() == 1$, 9.5\% of flaky tests were always flaky in their 35 previous builds.
Still, there is a non-negligible amount (45.3\%) of fault-revealing tests that were flaky at least once in previous builds considered: with a $flakeRate() > 0$.
From these observations, we may suggest that the \textsf{flakeRate()} can be used to detect flakiness.
Nevertheless, there is still an important overlap between the history of flaky tests and fault-revealing tests. 


\section{Objectives and Methodology}
\label{sec:evaluation}

\subsection{Research questions}

We start our analysis by assessing the effectiveness of the existing flakiness prediction methods in our project by considering the critical cases where fault-triggering test failures are flagged as flaky by the prediction methods in various CI cycles. We thus are interested in investigating the methods' performance under realistic settings, i.e., correctly detected and missed flaky and fault-triggering test failures, when trained with past CI data and evaluated on future ones. In contrast to previous work, this analysis introduces a new dimension in the evaluation of flakiness predictions which is the investigation of what we lose when adopting a prediction method (the fault-triggering failures classified as flaky). Therefore, we ask:

\begin{description}
\item[RQ1:] How well do flaky test prediction methods discern flaky test failures from fault-triggering ones? 
\end{description}

To establish realistic settings, we train the prediction models using the information available (flaky tests and non-flaky ones) at a given point in time, where we have sufficient historical data to train on. We then evaluate the models in subsequent commits with respect to flaky and fault-triggering test failures. 
We replicate the vocabulary-based methods since they are popular, easy to implement and quite effective, and aimed at learning to predict flaky tests, as done by previous studies.

After checking the prediction performance in a realistic setting (test failures), we repeat the entire process but now we train on historical test failures instead of tests. We make this adaptation with the hope of improving further our predictions and perhaps improving our understanding of the impact that such predictions may have on missed fault-triggering test failures (those marked by the models as flaky). Hence, we ask: 

\begin{description}
\item[RQ2:] How well do flaky test failure prediction methods discern flaky test failures from fault-triggering ones? 
\end{description}

Finally, we wish to be comprehensive, so we also optimize and extend the prediction methods with additional features, some of which were suggested by previous studies (the flake rate~\cite{Kowalczyk2020}, the run duration~\cite{alshammari2021flakeflagger}) and some dynamic features (test run status, test run tag status, test run duration) that we found by us when experimenting with the flaky tests. Thus, we ask:

\begin{description}
\item[RQ3:] Can we improve the accuracy of the flaky test failure predictions by considering dynamic test execution features?
\end{description}

To answer RQ3, we repeat the analysis carried out for RQ2 but now we are training and optimising for the additional features that we determined during our analysis and check the performance we achieve with regards to test failures, as performed in RQ2.



\subsection{Experimental procedure}

\subsubsection{Selection of a flaky test detection approach} 
\label{sec:evaluationSelection}
Being a recent topic of interest, several techniques have been introduced in the scientific literature. Approaches relying on code coverage such as FlakeFlagger~\cite{alshammari2021flakeflagger} or DeFlaker~\cite{Bell2018} are challenging to implement in our case. Chromium's code base consists of several languages and code coverage is both costly and non-trivial to retrieve. Test smells~\cite{camara2021use} approaches are also difficult to extract as tests are written in many different languages and tools do not always exist. 
Having those constraints in mind, we decide to use the vocabulary-based approach introduced by Pinto\etal ~\cite{Pinto2020}. It received significant attention with several replication studies conducted~\cite{Haben2021,Camara2021VocabExtendedReplication} and follow-up studies and its lightweight makes it easy to implement regardless of the languages being used.

\subsubsection{Training and validation of the existing approaches (RQ1)} 
We evaluate the ability of the vocabulary-based approach, trained to differentiate flaky from non-flaky tests and used to predict flaky test failures. To do so, we divide our dataset into a training set (containing test information about the first 8,000 builds) and a test set (containing test information about the last 2,000 builds). We train our model following the existing methodologies. The flaky set includes all tests marked as flaky in the training set. 
The non-flaky set includes all fault-revealing tests and all passing tests in the 8,000\textsuperscript{th} build (\ie the last build of the training set) minus the tests that are found as flaky in any of the builds under study (to increase the confidence of being non-flaky). The test set includes all flaky test failures and all fault-triggering failures (reported by fault-revealing tests). The test set is common in all RQs.

\subsubsection{Implementation of a failure classifier (RQ2 and RQ3)}
We select flaky failures in our dataset as all failures produced by flaky tests and fault-triggering failures are all failures produced by fault-revealing tests. There are no duplicated data in the case of test failures, as each test execution is unique. For RQ2, we train our classifier on non-flaky executions (passing and fault-revealing tests execution) and flaky failures. In RQ3, we report the performance of a model using execution features (run duration and flake rate).

\subsubsection{Time-sensitive evaluation}
\begin{figure*}[t]
\vspace{-1.4em}
\centering
\includegraphics[width=0.75\textwidth]{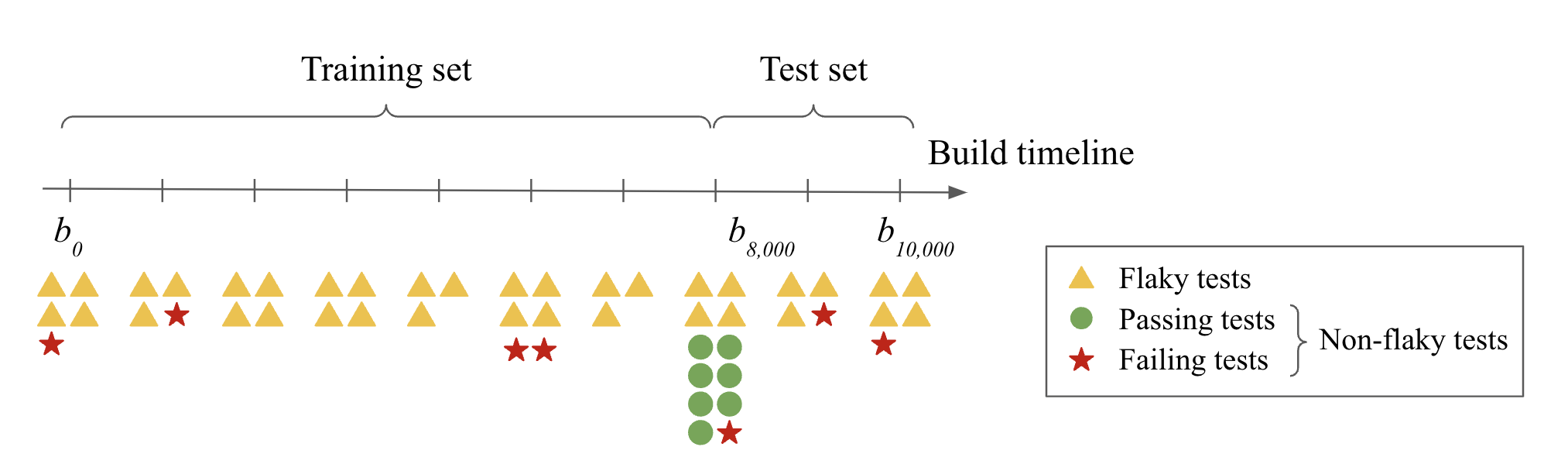}
\vspace{-1.1em}
\caption{The data collected from Chromium's CI consists of flaky, fault-revealing and passing tests spread across 10,000 builds. The build timeline ranges from build \textit{$b_0$} to \textit{$b_{10,000}$} and depicts the distribution of the collected tests: flaky tests are spread across all builds and fault-revealing tests happen occasionally. Due to a large number of passing tests, we collected them from the \textit{$b_{8,000}$} build (\ie at the end of our training set).}
\label{fig:dataset}
\vspace{-0.2em}
\end{figure*}

We split our data in two parts: the first 80\% builds are selected as a training set and the last 20\% as a holdout set. By doing so, we respect the evolution of failures across time and avoid any data leakage that could occur by randomly selecting data. This time-sensitive aspect is very important to consider. We found that not taking this condition into account and training a model on a shuffled dataset would greatly overestimate the performance. Figure~\ref{fig:dataset} shows a representation of our dataset. Flaky tests are present in all builds and Fault-revealing tests are occasional: they happen in \nicefrac{1}{4} of builds (See ~\ref{sec:discussion}). To make mitigate imbalance, we collected all passing tests for 1 build: $b_{8,000}$ and use them in our set of non-flaky tests, for training.

\subsubsection{Classifier selection and pipeline description}
We use random forest to perform the predictions. Unfortunately, our dataset is imbalanced, especially in the case of failure predictions where the minority class constitutes 1\% of the data. Using a simple random forest would greatly increase the chance of having few or no elements from our minority class in the different trees, making the overall model poor in predicting the class of interest. To alleviate this issue, we decide to use a Balanced Random Forest classifier\cite{chen2004using} to facilitate the learning. This implementation artificially modifies the class distribution in each tree so that they are equally represented.

To represent the test sources, we use CountVectorizer to convert texts as a matrix of token counts. This technique, known as bag-of-words, is used in previous vocabulary-based approaches~\cite{Pinto2020,Haben2021,Camara2021VocabExtendedReplication,Bertolino2020,olewickiBrown}. Test vectors initially contain as many features as there are words in all test sources. The overall generated dictionary can have a significant size. To reduce the number of features, we use feature selection. This step has several advantages. First, it removes irrelevant features which reduces noise in the data and thus, improves the overall performance of the model. Secondly, selecting only important features can help in model interpretability. Finally, as fewer features are used, it reduces training time. 

To this end, we use SelectKBest\cite{selectkbest} which retains the $k$ highest score features based on the univariate statistical test $\chi^2$.
Hyper-parameters of the machine learning pipeline, \ie the number of trees in the forests, the sampling strategy for SMOTE and the number of features to be retained are tuned using a grid search approach and cross-validation in the training set. Once optimized, we retrain a model fitted on the whole training set and evaluate it on the holdout set.

\subsubsection{Metrics}
Finally, to evaluate the different models, we rely on the following metrics derived from true positives (TP), true negatives (TN), false positives (FP) and false negatives (FN):
    \[
    \textbf{Precision} = \frac{TP}{TP+FP} \quad \quad \quad \textbf{Recall} = \frac{TP}{TP+FN}
    \]

The accuracy of a model is sensitive to class imbalance. In particular, the precision and recall metrics can easily be impacted when one class is underrepresented. To alleviate this issue, we report the Matthews Correlation Coefficient (MCC) which is a more reliable statistical rate to avoid over-optimistic results in the case of an imbalanced dataset \cite{chicco2020advantages}.
This metric takes into consideration all four entries of the confusion matrix. MCC ranges from -1 to 1 and is given by the following formula: 
    \[
    \textbf{MCC} = \frac{TN \times TP - FP \times FN}{\sqrt{(TN+FN)(TP+FP)(TN+FP)(FN+TP)}}
    \]

In addition to those metrics, we also report the false positive rate (FPR), that is, the ratio of fault-triggering failures misclassified as flaky over all fault-triggering failures. It is defined by:
    \[
    \textbf{FPR} = \frac{FP}{FP+TN}
    \]
    
\section{Results}
\label{sec:results}


\subsection{RQ1: Discerning flaky from fault triggering test failures when training on tests}

We trained a model on 69,159 passing tests, 910 fault-revealing tests and 8,857 flaky tests (unique tests).
Then, we evaluated it on 217,503 failures caused by flaky tests and 2,320 fault-triggering failures caused by fault-revealing tests. Table ~\ref{table:rq1} reports the obtained performance. Similar to the performance achieved by previous vocabulary-based models on other datasets, our model was able to reach high accuracy with a precision of 99.2\% and a recall of 98.9\%. However, we note a high false-positive rate. This is due to an important amount (76.2\%) of fault-triggering failures classified as flaky (FP). This is concerning: fault-triggering failures should not be misclassified as they reflect the existence of real bugs. Overall, the MCC value we have is 0.20, and shows that the model struggles (compared to random selection) to identify fault-triggering failures. 

\begin{table}[ht]
\caption{Vocabulary-based model performance for the prediction of \textit{flaky test failures vs fault-triggering failures} when trained on flaky vs non-flaky (fault-revealing and passing tests). Despite a high accuracy on flaky failures, the low MCC and high FPR show us that it remains challenging for the model to classify negative elements (in our case: fault-triggering failures)}
\vspace{-1.0em}
\label{table:rq1}
\begin{center}
\begin{tabular}{c|c|c|c} 
 \toprule
 \textbf{Precision} & \textbf{Recall} & \textbf{MCC} & \textbf{FPR} \\ [0.5ex] 
 \midrule
 99.2\% & 98.9\% & 0.20 & 76.2\% \\ 
 \bottomrule
\end{tabular}
\end{center}
\end{table}

\begin{figure}[!htbp]
\centering
\vspace{-1.2em}
\includegraphics[width=0.45\textwidth]{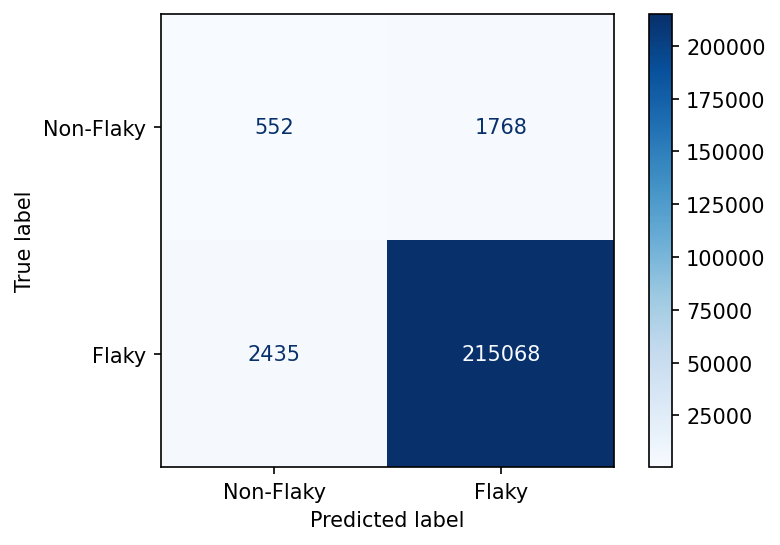}
\vspace{-1.3em}
\caption{Confusion matrix for the vocabulary-based model. High accuracy is reached similar to the performance reported in previous works. Nonetheless, 1,768 (76.2\%) out of the 2,320 fault-triggering failures are mislabeled as flaky.}
\label{fig:confMatrix}
\end{figure}

The confusion matrix of our model decisions is depicted in Figure ~\ref{fig:confMatrix}. The x-axis reports the predicted label and the y-axis the actual label. Correct classifications are displayed in the top left (TN) and bottom right (TP). We clearly observe that the model is able to detect flaky tests with high precision. We also see that 2,435 flaky tests are classified as non-flaky (FN). This number is also important to consider: it translates in all cases where developers will be required to investigate irrelevant failures.

We want to further understand the reasons behind the classification of fault-triggering failures. Therefore, we analyse the (fault-revealing) tests causing those failures. Out of the 2,320 fault-triggering failures, 1,768 are in the set of false positives (76.2\%) among which we found 463 (20\% of all fault-triggering failures) whose tests have a history of flakiness (flakeRate > 0) in the training set and 1,305 (56.2\% of all fault-triggering failures) without flakiness history. Here it must be noted that depending on the size of the history considered, we may have more tests with past flakiness or less. Overall in our data, 1/3 of all fault-triggering failures are due to tests that have exhibited flakiness behaviour.



\begin{tcolorbox}[
    left=2pt,right=2pt,top=2pt,bottom=2pt, 
    arc=0pt, 
    boxrule=1.2pt 
]
\textbf{RQ1:} Similar to previous studies, we report accurate predictions when aiming at flaky tests. However, a high proportion (76.2\%) of all fault-triggering failures is classified as flaky (missed faults) and still an important number (2,435) of flaky tests marked as fault-triggering failures (false alerts).
\end{tcolorbox}

\subsection{RQ2: Discerning flaky from fault triggering test failures when training on test failures}

The results from RQ1 show that a vocabulary-based model trained to detect flaky tests would still yield an important number of missed faults and false alerts despite having high accuracy. Thus, our goal with RQ2 is to check whether by training our vocabulary-based model we can improve the performance of recognising fault-triggering failures. 

Table ~\ref{table:rq2and3} reports the results of such a model. In particular the first row of reports results based on failure training while the second row reports results related to RQ3. Similar to RQ1, we see a high precision and recall, 99.7\% and 91.3\% respectively, when predicting flaky failures. More interestingly, the MCC slightly increased to 0.25.

\begin{tcolorbox}[
    left=2pt,right=2pt,top=2pt,bottom=2pt, 
    arc=0pt, 
    boxrule=1.2pt 
]
\textbf{RQ2:} When training on test failures, solely relying on test code vocabulary as features, to predict if a test failure is flaky or fault-triggering, model performance slightly improve but is still not effective in the context of the Chromium CI.
\end{tcolorbox}

\begin{table}[ht]
\caption{Vocabulary-based model performance for the prediction of \textit{flaky failures vs fault-triggering failures} when training on flaky vs non-flaky (fault-triggering and passing test executions). The approach does not work when solely relying on static features (\ie the test source code) and is improved when considering execution features.}
\label{table:rq2and3}
\begin{center}
\begin{tabular}{c|c|c|c|c} 
 \toprule
 \textbf{Execution features} & \textbf{Precision} & \textbf{Recall} & \textbf{MCC} & \textbf{FPR} \\ [0.5ex] 
 \midrule
 No & 99.7\% & 91.3\% & 0.25 & 20.3\%\\ 
 Yes & 99.5\% & 98.7\% & 0.42 & 42.3\%\\ 
 \bottomrule
\end{tabular}
\end{center}
\end{table}

\subsection{RQ3: Improving the accuracy of the flaky test failure predictions}

In this RQ we check he performance of the vocabulary-based models on the failure classification task when considering additional features from the test executions (run duration, and tests' historical flake rate). These features reflect better the characteristics of the test executions and are linked with test flakiness thereby leading to better results. 

In particular, the second row of Table~\ref{table:rq2and3} reports the related performance. We observe an improvement compared to the model that relies only on vocabulary (RQ2). This new model achieves a similar precision and recall of 99.5\% and 98.7\% and an improved MCC value 0.42, indicating a better performance in comparison to randomly picked selections. We see that the FPR increased to 42.3\%. Together, the results can be explained by fewer false alerts: flaky failures being marked as fault-triggering by the model. 

\begin{tcolorbox}[
    left=2pt,right=2pt,top=2pt,bottom=2pt, 
    arc=0pt, 
    boxrule=1.2pt 
]
\textbf{RQ3} When equipped with execution-related features, the vocabulary-based prediction methods do a better job of distinguishing flaky failures from fault-triggering failures (0.42 MCC). Still, the need remains for dedicated methods to successfully learn this challenging classification task.

\end{tcolorbox}
\section{Discussion}
\label{sec:discussion}

We seek to better understand the results showing that existing approaches targeting the detection of flaky tests missed a non-negligible part (76.2\%) of fault-triggering failures by classifying them as flaky. To do so, we investigate the following aspects regarding \textit{the entire dataset}.

We first report general information about the prevalence of flaky tests and fault-revealing tests in order to have a better representation of failures occurring in each build. Then, we report the number of fault-revealing tests also found as flaky by the Chromium CI (reruns) in other builds (we further refer to them as fault-revealing flaky tests). Finally, we also check for the number of failing builds that only contain fault-revealing flaky tests. We consider those builds as particularly harmful, as regressions are not reported by any reliable tests.

\begin{figure}[!htbp]
\centering
\includegraphics[width=0.47\textwidth]{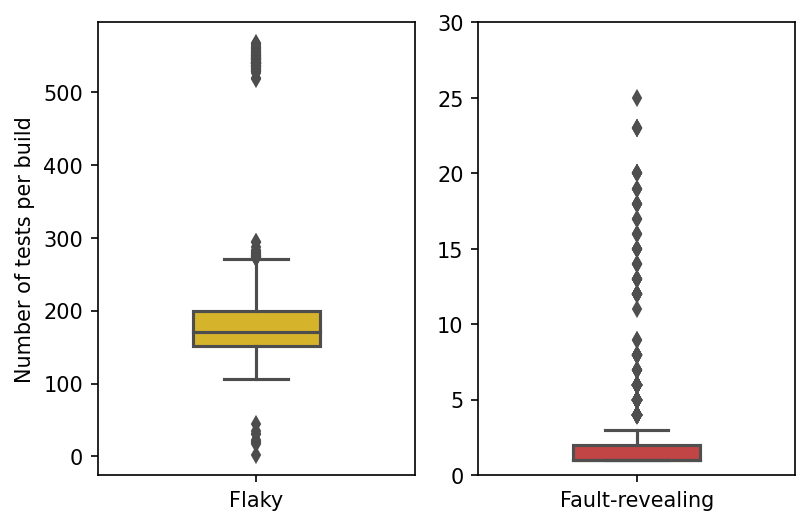}
\caption{Number of flaky tests and fault-revealing tests per build. On average, there are 250 flaky tests per build and 1 fault-revealing test per failing build.}
\label{fig:testsPerBuild}
\end{figure}

Figure~\ref{fig:testsPerBuild} shows the distribution of flaky tests and fault-revealing tests in the studied builds. We observe that there is an average of 178 flaky tests per build with a low standard deviation (41), showing that flakiness is prevalent in the Chromium CI. In the case of fault-revealing tests, taking into account all builds would result in an average number of tests close to 0 as a majority of builds are exempt from them. Thus, for better visualisation, we only considered builds containing at least one fault-revealing test (\ie failing builds). The average number of fault-revealing tests per failing build is 2.7.
The standard deviation for fault-revealing tests is 14.9 and the number of fault-revealing tests reported in one build goes up to 579 in our dataset.

\begin{table}[ht]
\caption{Number of builds containing each studied test type. All builds contain flaky tests. \nicefrac{1}{4} contain fault-revealing tests. Among the failing builds, \nicefrac{3}{4} contain only fault-revealing tests that are flaky in other builds.}
\label{table:DiscBuilds}
\begin{center}
\begin{tabular}{c|c} 
 \toprule
 \textbf{Builds containing} & \textbf{Number} \\ [0.5ex] 
 \midrule
 Flaky tests & 10,000 \\ 
 Fault-revealing tests & 2,415 \\ 
 Fault-revealing flaky tests & 1,974 \\ 
 Exclusively fault-revealing flaky tests & 1,766 \\ 
 \bottomrule
\end{tabular}
\end{center}
\end{table}

Table~\ref{table:DiscBuilds} provides, for each type of test, the number of builds that contain at least one instance of this type. We notice that all builds contain at least one flaky test (a test that flaked during this build). In Chromium CI, flaky tests are non-blocking and will not cause a build failure. That is, tests flaking within the build are ignored during this build. Developers are expected to investigate test failures only when they occur consistently across 5 reruns (resulting in a fault-revealing test). Such fault-revealing tests occur in 24.15\% of the builds (i.e. in 2,415 builds). Interestingly, 1,974 of these builds (i.e. 81.73\%) contain fault-revealing tests that flaked in previous builds, indicating that \emph{tests with a flake history should not be ignored in future builds}. Perhaps worse, in 1,766 builds \emph{all} fault-revealing tests have flaked in some previous builds, indicating that no "reliable" tests identified the fault(s).


\begin{figure}[!htbp]
\centering
\includegraphics[width=0.5\textwidth]{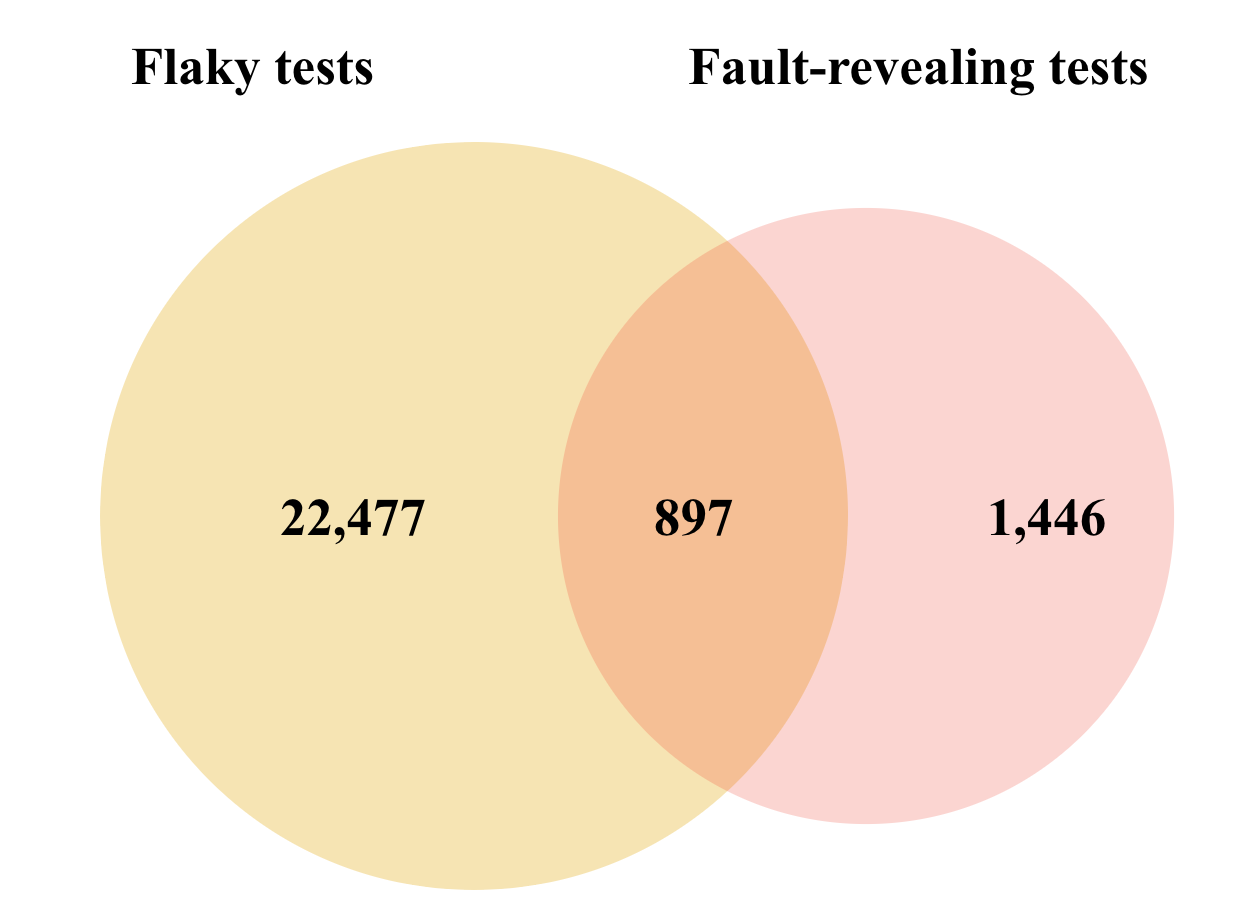}
\caption{Distribution of tests in our dataset. 22,477 tests are exclusively flaky among all builds. 2,343 tests are fault-revealing, among which \nicefrac{1}{3} are flaky in other builds.}
\label{fig:venn}
\end{figure}

We now investigate the status of all tests across all builds -- see Figure~\ref{fig:venn}. Among the 209,530 tests of the Chromium project, 24,820 have failed in at least one build, including 22,477 that were always flaky. Thus, 2,343 tests were fault-revealing in at least one build, i.e., they attested the presence of faults. 897 of these tests were also detected as flaky in at least one other build. That is, \emph{38.3\% of tests that have been useful to detect regressions have also a history of flakiness}.


\begin{tcolorbox}[
    left=2pt,right=2pt,top=2pt,bottom=2pt, 
    arc=0pt, 
    boxrule=1.2pt 
]
Flakiness affects all Chromium CI builds and mixes critical (fault-revealing) signals with false (flakiness) signals. Indeed, 81.7\% of builds contain fault-revealing tests that were flaky in some previous builds, and 38.3\% of all tests flake in some builds and reveal faults in other builds. 
\end{tcolorbox}
\section{Threats to validity}
\label{sec:threats}

\subsection{Construct validity}
The main threat to the construct validity of our study lies in our use of vocabulary-based approaches as predictors for flaky tests and flaky failures. Approaches leveraging other features, \ie dynamic, static, or both, could perform better. As explained in section ~\ref{sec:evaluationSelection}, many features are difficult to extract in the case of Chromium (\eg test smells, test dependency graphs...) and features relying on code coverage are not considered due to the overheads they introduce, perhaps also because it is challenging to instrument the entire codebase. Although the absence of coverage features limits our features, the same situation appears in many major companies such as Google and Facebook. Nevertheless, our key insight is that many regression faults are discovered by flaky tests, meaning that they would have been missed even by any flaky test detector that correctly considers them as flaky. 

\subsection{External validity}
With our study, we show that detecting flaky tests in the CI with the goal of disregarding their signals is harmful as it can miss reporting many regressions. This is the case for the Chromium project and, while we believe Chromium to be representative of other large software systems, we cannot guarantee that findings would generalise to other projects. Similarly, the performance of the different models we report may vary depending on the project. Here, we mainly focus on web/GUI tests and flakiness might have different causes in HTML and Javascript testing compared to other programming languages. Nevertheless, we believe that in other cases flaky tests are useful since developers tend to keep flaky tests instead of discarding them. Therefore, flaky test signals should always be considered with caution. 

\subsection{Internal validity}
We assume that all fault-revealing tests in our dataset indeed reflect one or several regressions in the code. This is the information reported by the Chromium CI as of today. It is possible that some fault-revealing tests are actually flaky tests that were not reruns enough time. However, no amount of reruns can guarantee that a test is not flaky. As this information is currently used by Chromium's developers and further verification is non-trivial, we rely on it as ground truth for our dataset. Passing tests used as non-flaky tests could also be prone to mislabel in our dataset. However, there is a consequent number of passing tests and it is unlikely that many would actually be flaky. Furthermore, to strengthen our confidence in our set of non-flaky tests, we remove from the set of passing tests any tests that were found to be either flaky or fault-revealing in any of the 10,000 builds.

Additionally, it is possible that more than one regression fault is present in the case of several fault-revealing tests failing in one build. Although this could alter the results we report in RQ1, it would actually strengthen our key message as even more regression faults could be missed. 
\section{Conclusion}
\label{sec:conclusion}

In this paper, we investigated the utility of existing flaky test prediction methods in the context of a continuous integration pipeline. To do so, we collected
information about 23,374 flaky tests and 2,343 fault-revealing tests totalling a dataset of 1.8 million test failures representing the actual development process over 10,000 builds in a period of 9 months. We evaluated prediction methods when asked to classify flaky failures and fault-triggering failures. We reported similar performance compared to previous works in terms of precision and recall. Despite the (very) high accuracy to detect flaky failures, we were surprised to find that 76.2\% of fault-triggering failures were misclassified as flaky by the prediction methods. To explain this behaviour, we showed that flaky tests have in fact a strong ability to detect faults: \nicefrac{1}{3} of faults are revealed by tests that were found to be flaky at some point.

These findings motivate the need for failure-focused prediction methods. Thus, we continued our analysis by checking the performance of failure-focused models (trained on failures instead of tests). We found that they result in similar accuracy and fewer false positives. We also found that considering test execution features such as the run duration and the historical flake rate was helpful to increase its ability to discern flaky failures and fault-triggering failures. Still, we believe that the current performance is not actionable in Chromium's CI pipeline. Further research should focus on the vastly-ignored problem of predicting failures instead of tests.

 Our future agenda aims at improving the performance of flaky test failure detection techniques by using additional features and artificial data (augmenting the training data with new positive examples to tackle the class imbalance issues). We also plan to develop failure-cause interpretations for the techniques so that they could be usable by the Chromium developers.  \\
 \\


\bibliographystyle{ACM-Reference-Format}
\bibliography{main}

\end{document}